\documentclass[10pt, conference,letterpaper]{IEEEtran}
\usepackage[utf8]{inputenc}
\usepackage[T1]{fontenc}
\usepackage{lmodern} 
\usepackage{cite}

\ifCLASSINFOpdf
  \usepackage[pdftex,final]{graphicx}
  \graphicspath{{../pdf/}{../jpeg/}}
  \DeclareGraphicsExtensions{.pdf,.jpeg,.png}
\else
  \usepackage[dvips]{graphicx}
  \graphicspath{{../eps/}}
  \DeclareGraphicsExtensions{.eps}
\fi

\usepackage{tabularx}
\usepackage{caption}
\usepackage{subcaption}
\usepackage[cmex10]{amsmath}
\usepackage{amsfonts}
\usepackage{algorithmic}
\usepackage{algorithm}

\usepackage{url}
\usepackage{hyperref}
\usepackage{hhline}
\usepackage{mathtools}
\usepackage{booktabs} 

\begin{document}
%
\title{Scalable BGP Prefix Selection for Effective Inter-domain Traffic Engineering}

\author{
\IEEEauthorblockN{Wenqin Shao, Luigi Iannone \\and Jean-Louis Rougier}
\IEEEauthorblockA{Telecom ParisTech\\
{\footnotesize \{firstname.lastname\}@telecom-paristech.fr}}
\and
\IEEEauthorblockN{Fran\c{c}ois Devienne \\and Mateusz Viste}
\IEEEauthorblockA{Border 6\\
{\footnotesize  \{firstname.lastname\}@border6.com}}
}

\maketitle
\thispagestyle{plain}
\pagestyle{plain}

\begin{abstract}
Inter-domain Traffic Engineering for multi-homed networks faces a scalability challenge, as the size of BGP routing table continue to grow.
In this context,
the choice of the best path must be made potentially for each destination prefix, requiring all available paths to be characterized (e.g., through measurements) and compared with each other.
Fortunately, it is well-known that a few number of prefixes carry the larger part of the traffic. 
As a natural consequence, to engineer large volume of traffic only few prefixes need to be managed.
Yet, traffic characteristics of a given prefix can greatly vary over time,
and little is known on the dynamism of traffic at this aggregation level,
including predicting the set of the most significant prefixes in the near future. 
Sophisticated prediction methods won't scale in such context.

In this paper, we study the relationship between prefix volume, stability, and predictability, based on recent traffic traces from nine different networks.
Three simple and resource-efficient methods to select the prefixes associated with the most important foreseeable traffic volume are then proposed.
Such proposed methods
allow to select 
sets of prefixes with both excellent representativeness (volume coverage) and stability in time, for which the best routes are identified.
The analysis carried out confirm the
potential benefits of a route decision engine. 

\end{abstract}

\IEEEpeerreviewmaketitle

\section{Introduction}

BGP (Border Gateway Protocl) multi-homing has been a long-time common practice for a variety of large networks, such as enterprise networks,  CDN (Content Delivery Networks), ISPs (Internet Service Provider), etc.\footnote{NAT-based multi-homing solutions, which do not require BGP, usually intended for smaller networks, are out of the scope of this paper.}
This solution is generally advocated for bringing enhanced connection reliability 
and improved traffic performance.

\begin{figure}[!tb]
\centering
\includegraphics[width=0.45\textwidth]{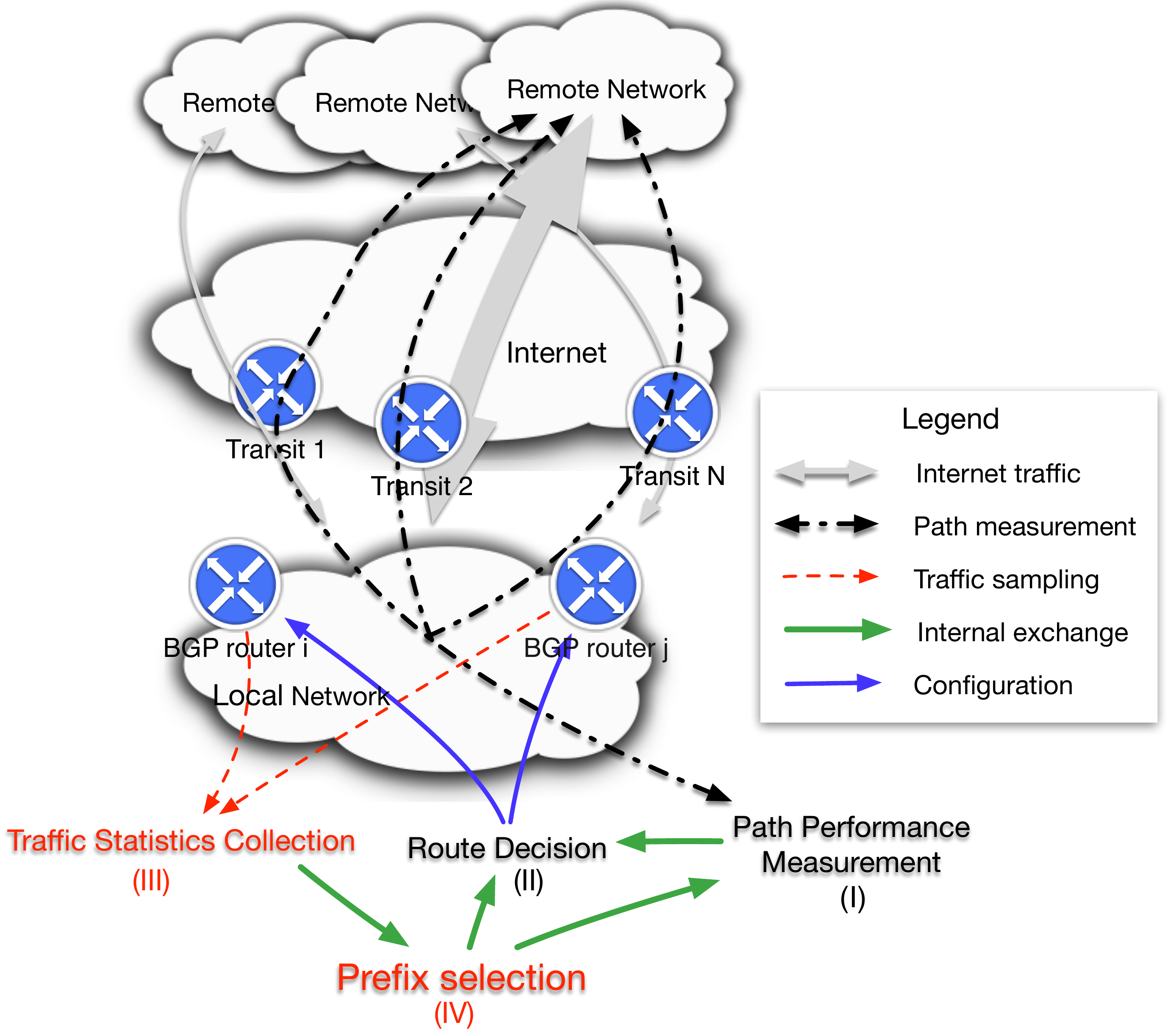}
\caption{Traffic Engineering with prefix selection.}
\label{fig:archi}
\end{figure}

Yet, this introduces the task of actually taking advantage of the multiple paths that become available,
%
%
when the network is connected to the Internet via multiple transit providers.
In such context, one or more BGP border routers are needed for the connection with transit providers and also used for Traffic Engineering (TE) purposes.
The best current practice in this context is to choose a best egress transit, in terms of certain optimization metrics, for each remote network (i.e., BGP prefix).
These metrics represent path performance information, (e.g., Round-Trip Time, packet loss, etc), which are not part of path attributes BGP advertises.
Fig.~\ref{fig:archi} provides a global picture of the tasks involved in performing TE in a multi-homed network. 
 %
%
%
Several OAM (Operations, Administration, and Maintenance) function blocks are needed (cf.,~Fig.~\ref{fig:archi}): (i) the performance of the paths through the different transits is measured; (ii) measures are then used by the ``route decision engine'', which selects the best egress transit for each BGP prefix and configure the BGP routers accordingly.
One major challenge of these two functions is to cope with the growing Internet Routing Information base (RIB)~\cite{potaroo}.
However, it is well-known that most traffic is generally associated with only a fraction of all the possible BGP prefixes (\cite{Fang1999, Feamster2003, Papagiannaki2005, Sarrar2012}), 
making it possible to focus only on such important prefixes. 
%

In this work, we propose a framework to improve the scalability of fine-grained inter-domain TE in the context of BGP, by concentrating on the most important BGP prefixes. 
This is achieved by predicting which prefixes will correspond to the most important traffic volumes in the near future.  
To this end, two additional function blocks are needed (cf., Fig.~\ref{fig:archi}): (iii) traffic analytics collection, which collects volume statistics; (iv) prefix selection process, which selects the set of the most important prefixes (i.e., the ones with the highest volume in the foreseeable future) and communicates the selected set to measurement and route decision function blocks.

Even if multi-homing has been used for more than a decade, there are still many challenges in fulfilling this goal.
Well established Time Series Forecasting (TSF) models and artificial neural networks (ANN) have been used previously in traffic prediction \cite{Papagiannaki2005, Cortez2006, Otoshi2013}.
These work targeted on highly aggregated inter-PoP (Point of Presence) traffic for off-line tasks, such as dimensioning.
These models are 
computationally heavy, 
and require pre-treatment and tuning to fit well to each individual trace, which makes them less applicable in the context of inter-domain TE, where more than $10^5$ prefixes exist. Therefore, less complex prediction methods 
are needed. 
Furthermore, the dynamism of traffic variation makes prediction even more difficult. Traffic dynamism has been studied at a per-flow level. For instance, it has  been shown by Walleriche et al.~\cite{Wallerich2006} that the bandwidth ranking of a 5-tuple flow can change drastically from one moment to another. It is however difficult to draw conclusions 
for the case of aggregated traffic (per BGP prefix). 
To our best knowledge, no study has given an in-depth investigation concerning the evolution in time of the traffic volume associated with BGP prefixes.

It is interesting to note that our work is similar to FIB (Forwarding Information Base) caching, which aims at installing only a small part of the routes, due to some software or hardware limitations. This problem has been extensively studied~(\cite{Iannone2007, Ballani2009, Kim2009, Zhang2012, Sarrar2012, Liu2015}).
 We have compared our prefix selection methods to these works 
 . We also believe that our study could be used for this problem as well. 

The main contributions of this work are:
\begin{itemize}
\item A study of real  traffic traces from networks of diverse profiles (ISPs, content providers and hosting providers) and from networks located in different countries (France, Germany, Poland, Spain, UK and USA).
\item A through investigation on the Internet traffic dynamism across BGP prefixes was conducted on these traffic traces, which demonstrates diverse dynamism characteristics.
\item The proposition of three scalable prediction methods. 
These methods are evaluated and compared to previous work. Our results show that the proposed schemes out-perform previous techniques 
both in term of performance and scalability. 
\item Finally, real measurements of the end-to-end path performance through each distinct transit provider for a set of prefixes is provided. The results confirmed older studies on multihoming in the fact that there could be significant differences among transit providers --- highlighting, therefore, that the potential gain of inter-domain TE in the context of a multi-homed site can be substantial. 
\end{itemize}

The rest of the article is organized as follows: in Sec.~\ref{sec:chara}, an in-depth study on the dynamism of traffic associated with BGP prefixes is given, based on which, three selection metrics are proposed and evaluated in Sec.~\ref{sec:sele}; Sec.~\ref{sec:rtt} evaluates the performance of different transit providers perceived by prefixes predictively selected and simulates the performance gain of a dynamic route selection algorithm based on latest RTT (Round-Trip Time) measurements toward selected prefixes; Sec.~\ref{sec:bg} presents previous related works; finally, we conclude our results and highlight the future directions in Sec.~\ref{sec:fut}.

\section{Characters of Internet traffic over BGP prefixes}
\label{sec:chara}

\begin{figure*}[!th]
\centering
        \begin{subfigure}[b]{0.49\textwidth}
				\centering
                \includegraphics[width=0.6\columnwidth]{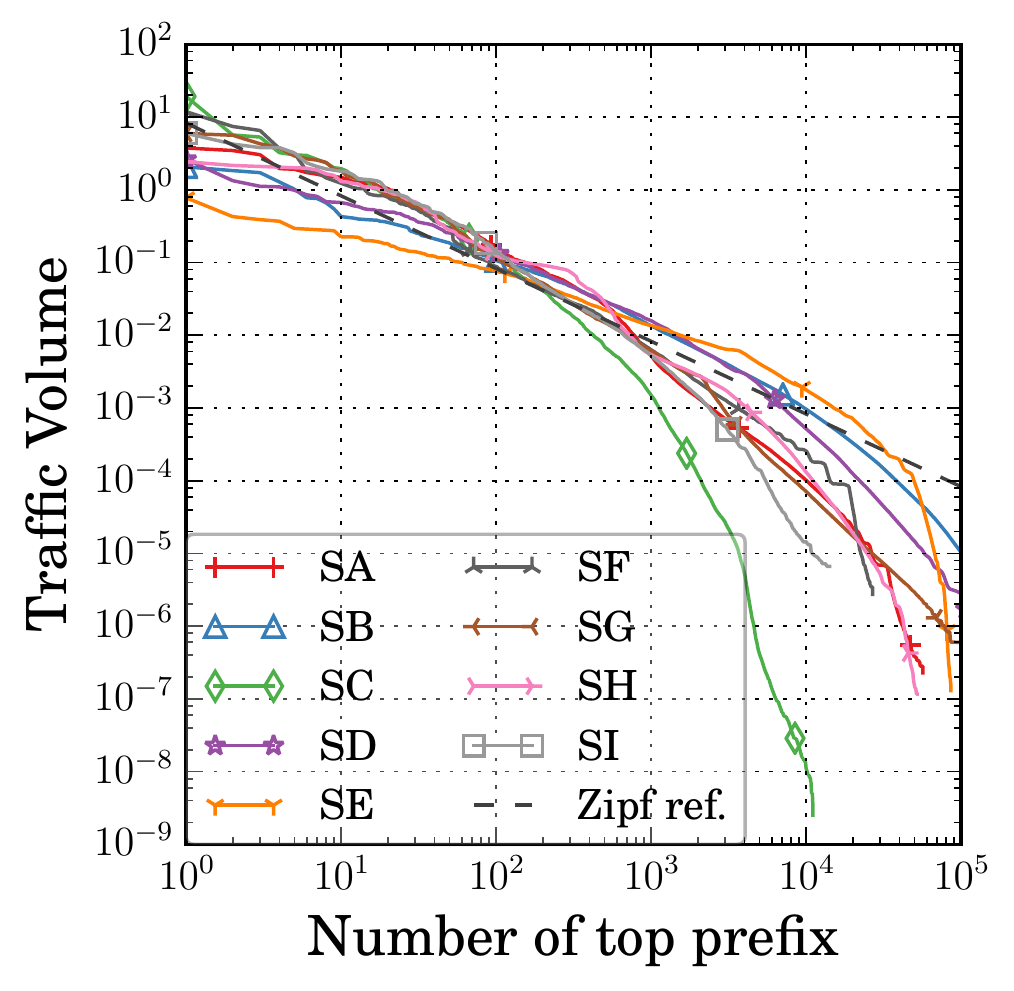}
                \caption{Week volume share, log-log.}
                \label{fig:week_zipf}
        \end{subfigure}  
        \hfill
        \begin{subfigure}[b]{0.49\textwidth}
        		\centering
                \includegraphics[width=0.6\columnwidth]{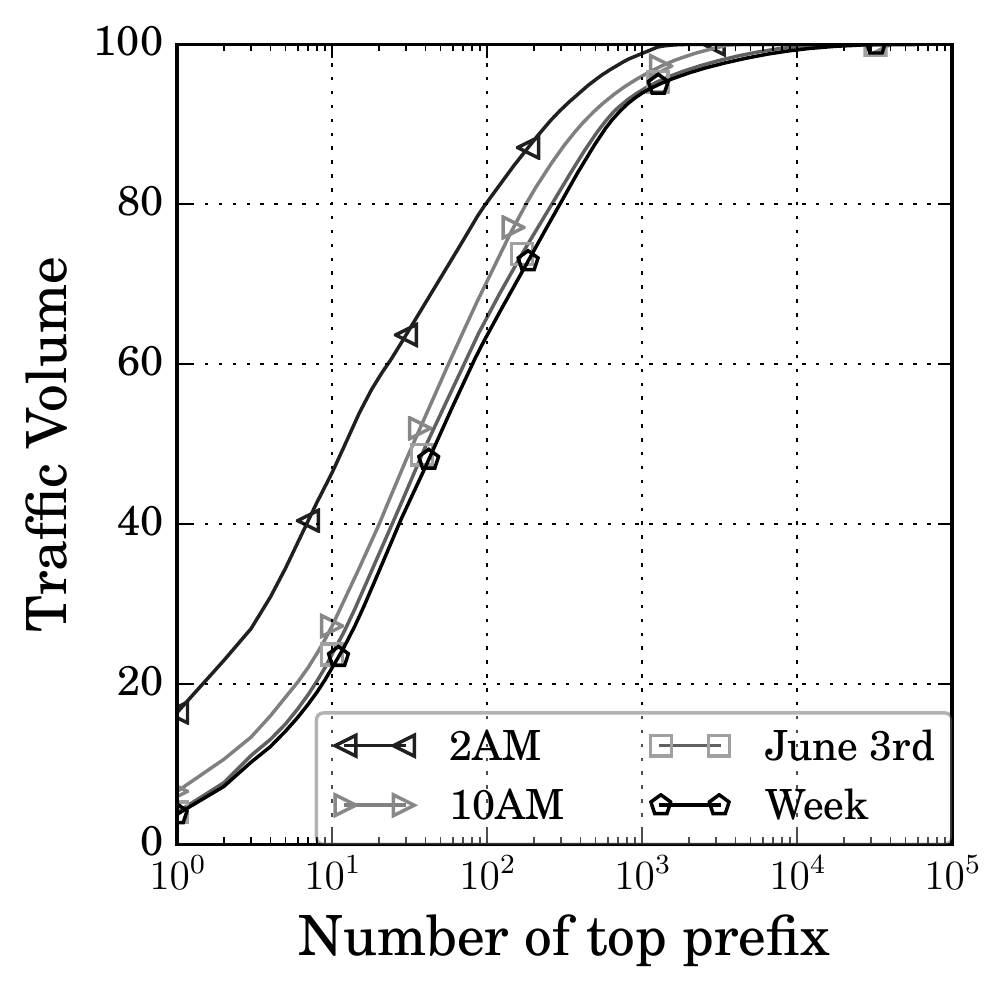}
                \caption{CDF different time spans, SA.}
                \label{fig:sa_cdf_multi_time}
        \end{subfigure}    
\caption{Traffic distribution among BGP prefixes.}\label{fig:traffic_dis_site}
\end{figure*}

\begin{table}[!tb]
\centering
\setlength{\tabcolsep}{0.5em}
\begin{tabular}{cccccccccc}
\toprule
 & \textbf{SA} & \textbf{SB} & \textbf{SC} & \textbf{SD} & \textbf{SE} & \textbf{SF} & \textbf{SG} & \textbf{SH} & \textbf{SI} \\
\midrule
\textbf{Type} & CP & ISP & HP & HP & CP & CP & ISP & CP & HP \\
\textbf{Vol.} & 133 & 528 & 6.7 & 1129 & 1871 & 5.1 & 0.2 & 29.9 & 6.2 \\
\bottomrule
\end{tabular}
\caption{Average traffic volume per hour (in GB) for the different measured networks}
\label{tab:network_type}
\end{table}

We base our study on working traffic traces collected from 9 networks of very different profiles listed in Table~\ref{tab:network_type}. 
They are either Content Providers (CP), Internet Service Providers (ISP) or Hosting Providers (HP). 
Traffic trace from each network covers a time period of two entire weeks, from May 25th, 2015 to June 8th, 2015, except for SB and SD, for which the traces only cover the second week (starting  June 1st, 2015).
These traces were sampled from real traffic, similarly to previous works on FIB caching \cite{Kim2009, Zhang2012}.
It has been shown that the bias introduced by sampling is negligible in this kind of use cases.


Traffic volume  for each prefixes are accumulated over 1 hour period.
In comparison, studies concerning FIB caching \cite{Sarrar2012, Zhang2012} use shorter time bins, from 1 second to 10 minutes, to capture instant changes of traffic variation.
In our case, we argue that 1 hour is an appropriate update interval for 
prefix selection, as we are more interested in long term evolution.
Changing the set of popular prefixes too frequently will make probing more difficult and might lead to frequent BGP route changes. 

\subsection{Traffic distribution over BGP prefixes}
\label{sec:dis}
As outlined earlier, this work is based on the fundamental assumption that traffic is concentrated on some popular traffic, as some previous work have shown \cite{Fang1999,Feamster2003, Wallerich2006}. For the sake of reliability, we quickly verify that our traffic traces demonstrate this property of concentration. 

In Fig.~\ref{fig:week_zipf}, the week volume share of each BGP prefix is represented. 
Prefixes are decreasingly sorted along the X-axis according to their cumulative volume fraction over the week.
We observe that the weekly volume associated with BGP prefixes can be approximately described by a reference Zipf's distribution with $N=10^5, s=1$ (dashed line)\footnote{Zipf's law defines that the $k^{th}$ most popular element among total $N$ elements has an occurrence share of $f(k,s,N)=\frac{1/k^s}{\sum_{n=1}^{N}1/n^s}$.}. 
Fig.~\ref{fig:sa_cdf_multi_time} compares the traffic distribution of SA over different time spans: volumes cumulated over one hour, at 2AM and 10AM, traffic cumulated over 24 hours (on June~3rd) and during the full week.

As expected, these graphs confirm that Internet traffic is highly concentrated on a few prefixes over multiple different periods of time, even after years of rapid RIB increase.
However, Fig.~\ref{fig:sa_cdf_multi_time} demonstrates that the level of traffic concentration over BGP prefixes is not stable but rather varies within a day, which leads to the study in the following section on traffic dynamism.

\subsection{Dynamism of traffic over BGP prefixes}
\label{sec:dyna}
In this section, we investigate the dynamism of traffic over BGP prefixes. 

For each prefix $P$ ever active during a week, it has a time series $v(P)={\left\{ v(P)_h\right\} }_{h=1, \dots, 168}$ that
stores its traffic volume each hour over the week. 
We define the Coefficient of Variation ($c_v$) of prefix $P$'s hourly volume series over the week as
$c_v(P) = \frac{\delta(v(P))}{\mu(v(P))}$ ,
where $\delta$ denotes the standard deviation of the hourly volume series and $\mu$ is the mean hourly volume over the week.
A large $c_v(P)$ value\footnote{By construction, the maximum $c_v$ for a hourly volume series of 168 in length is $\sqrt{167}$, corresponding to the case where the prefix in question is active during only one single hour throughout the entire week.} indicates  a large range of variation of traffic volumes w.r.t. its weekly average, and consequently more difficulty to anticipate the traffic volumes for this prefix $P$ \cite{He2005}.
\begin{figure}[!tb]
\centering
\includegraphics[width=0.48\textwidth]{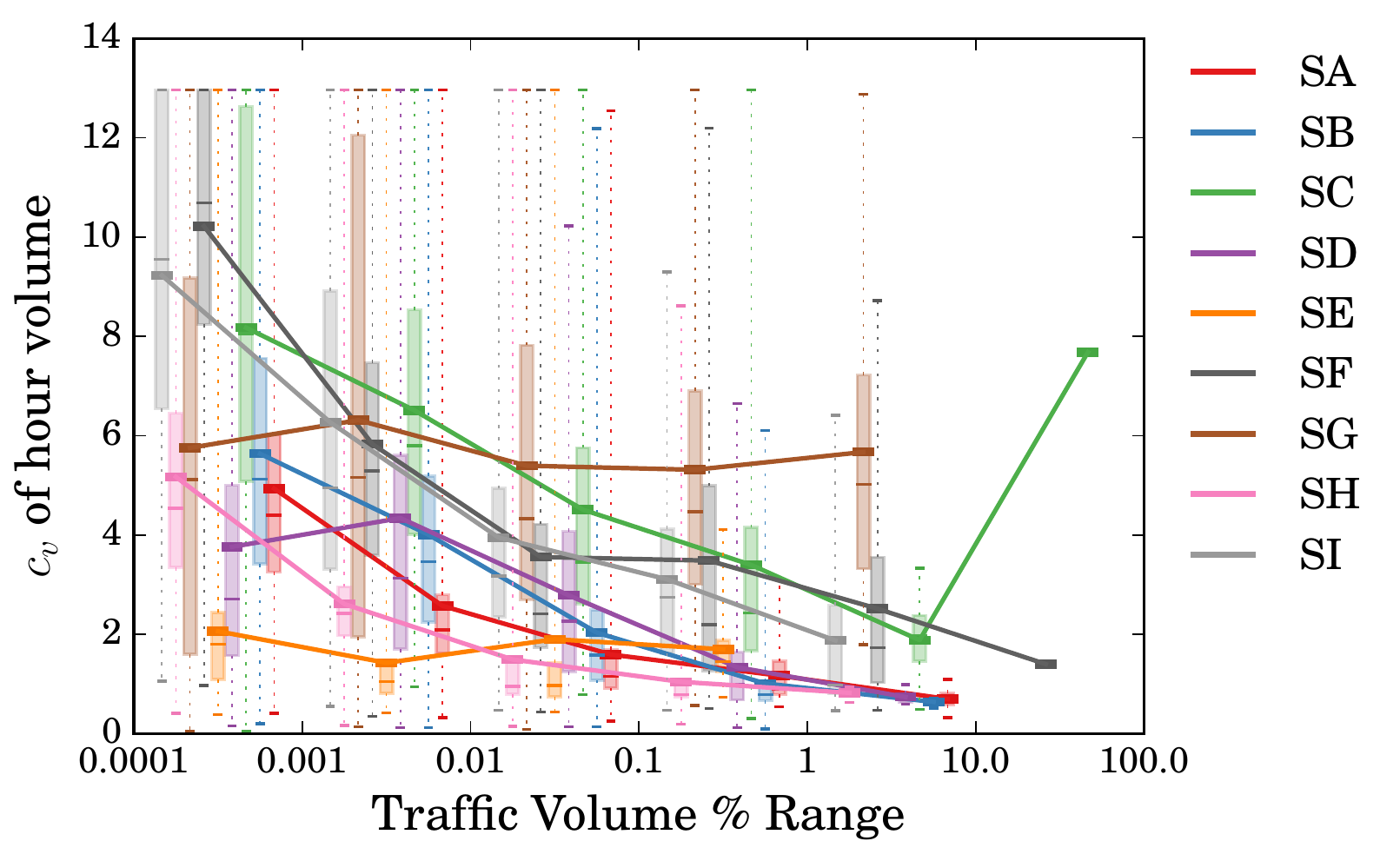}
\caption{Relation between $\{c_v(P)\}_P$  and  fraction of weekly volume for all BGP prefixes $P$. 
}
\label{fig:cv}
\end{figure}

Fig.~\ref{fig:cv} depicts the relation between the coefficient of variation and the ``importance'' of the prefix over a week. For each prefix $P$,  the X-axis corresponds to its fraction of volume over the week, grouped in six bins: $[10^{-4}, 10^{-3})$, $\dots$, to $[10,100)$. In the Y-axis, we use a statistical representation, with mean, median, 25-75 percentile of $\{c_v(P)\}_P$ (over all prefixes $P$ within the bin).
We observe that, for all the networks except SG and SE, $c_v$ of prefixes that contribute to the biggest amount of traffic tends to be constrained in a narrower range around a smaller mean value. On the contrary, the prefixes which correspond to smaller weekly volume tend to have larger coefficient of variation. 
This is an interesting property, as we can capture a significant part of the overall traffic (represented by these stable prefixes) by picking the best prefixes based on their average hourly volume. 


Next, we describe traffic dynamism from another perspective:
For each hour $h$, we define the $core_h$ as the set containing top prefixes that represent $95\%$ of total traffic. 
Table~\ref{tab:core_size} lists the average number of prefixes included in the \textit{core} and its percentage with regard to the average number of active prefixes each hour. 

\begin{table}[!tb]
\centering
\begin{tabular}{cccc}\toprule
\textbf{Name} & \textbf{Avg. prefix \#} & \textbf{Avg. prefix $\%$} & \textbf{Max prefix \#}\\
\midrule
SA & 629  & 17.87  & 1051\\
SB & 5264 & 9.45  & 13934\\
SC & 73  & 4.59    & 177\\
SD & 2481  & 17.35 & 3757\\
SE & 15501  & 53.61 & 20900\\
SF & 377  & 30.73    & 772\\
SG & 570 & 7.76    & 1766\\
SH & 965  & 19.42   & 1731\\
SI & 175  & 21.00    & 415\\
\bottomrule
\end{tabular}
\caption{Core statistics.}
\label{tab:core_size}
\end{table}

If a prefix is inside the \textit{core} at a certain hour, it can be regarded important for bringing a significant amount of traffic.
We thus defined the "\textit{core} presence" for a prefix $P$ at each hour $h$ as:
$cp(P)_h = 1 $ 
        when $P \in$ $core_h$,
         and 0 otherwise. 
The \textit{core} presence intensity over one week $I_{cp}(P, 168)$ is then defined as 
$I_{cp}(P,168) = \frac{1}{168} \sum_{i=1}^{168} cp(P)_i$.

\begin{figure}[!tb]
\centering
\includegraphics[width=0.48\textwidth]{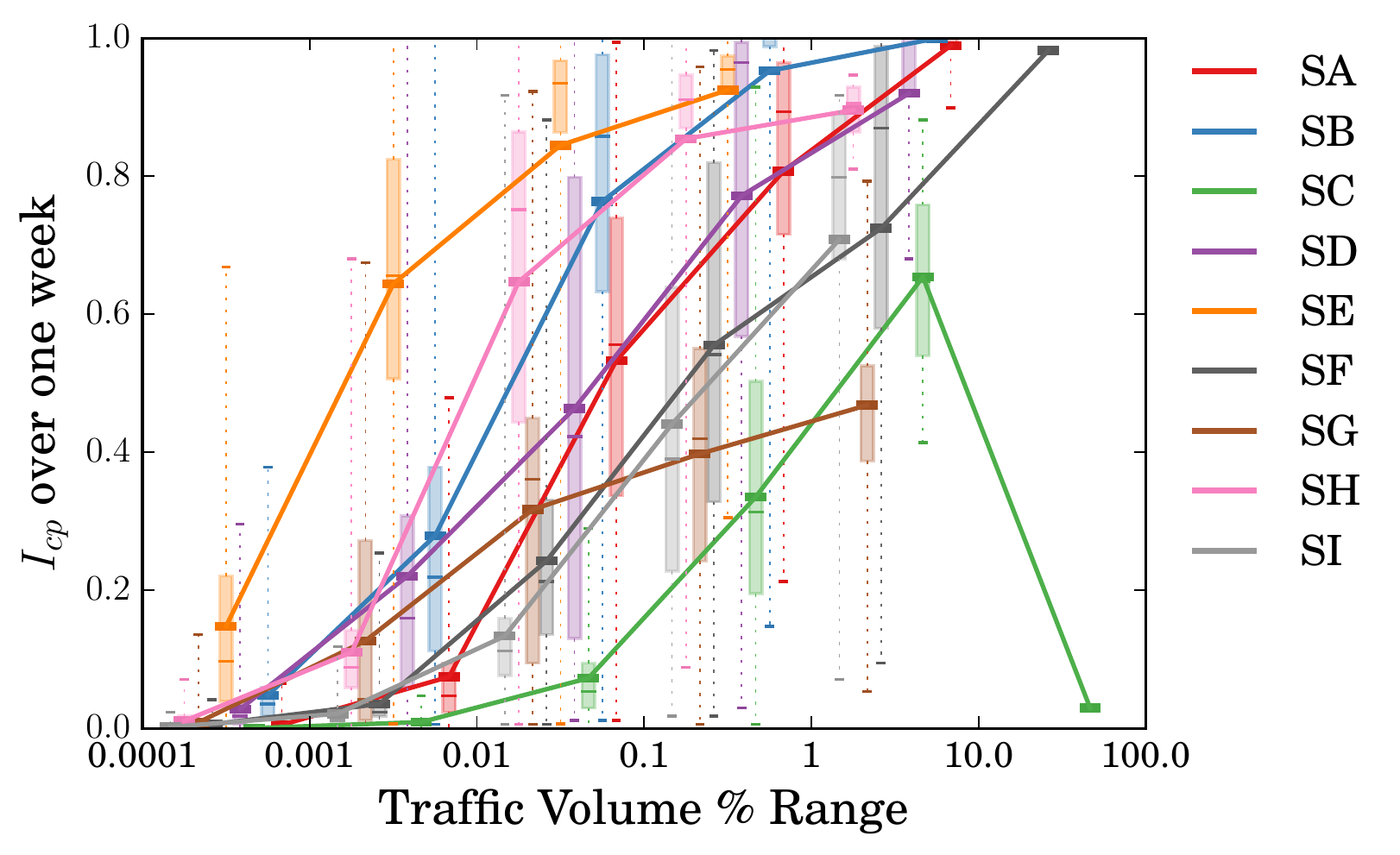}
\caption{Relation between $I_{cp}$ over the week and week volume fraction of BGP prefixes. 
}
\label{fig:cpi}
\end{figure}

Fig.~\ref{fig:cpi} plots the $I_{cp}$ over one week of each prefix (Y-axis) against its weekly volume fraction (in X-axis), using the same presentation as Fig.~\ref{fig:cv} described above.
For all the networks, we can see that prefixes with bigger week volume share are less likely to have a low $I_{cp}$ over the week, i.e. they appear frequently in the \textit{core}.
We can conclude that, by focusing on prefixes that intensely appear in the \textit{core} through the week, we will be able to capture a large part of the prefixes associated with important traffic volume over the week, and thus cover a large part of the overall traffic.

\begin{figure*}[!tb]
\centering
		\centering
        \begin{subfigure}[b]{0.49\textwidth}
        \centering
                \includegraphics[width=0.6\textwidth]{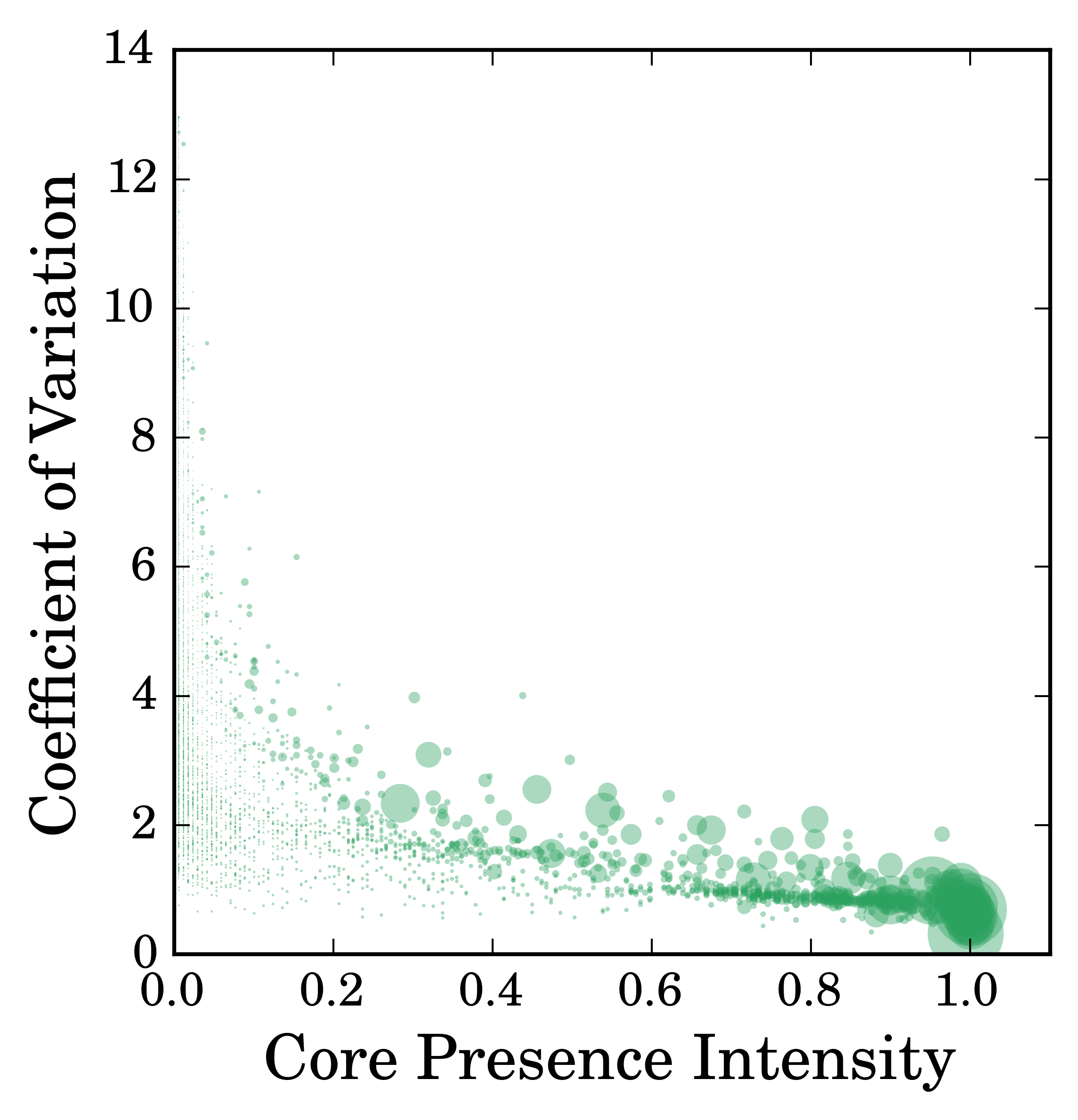}
                \caption{SA, 56684 prefixes}
                \label{fig:cv_cp_sa}
        \end{subfigure}
        \hfill
        \begin{subfigure}[b]{0.49\textwidth}
        \centering
                \includegraphics[width=0.6\textwidth]{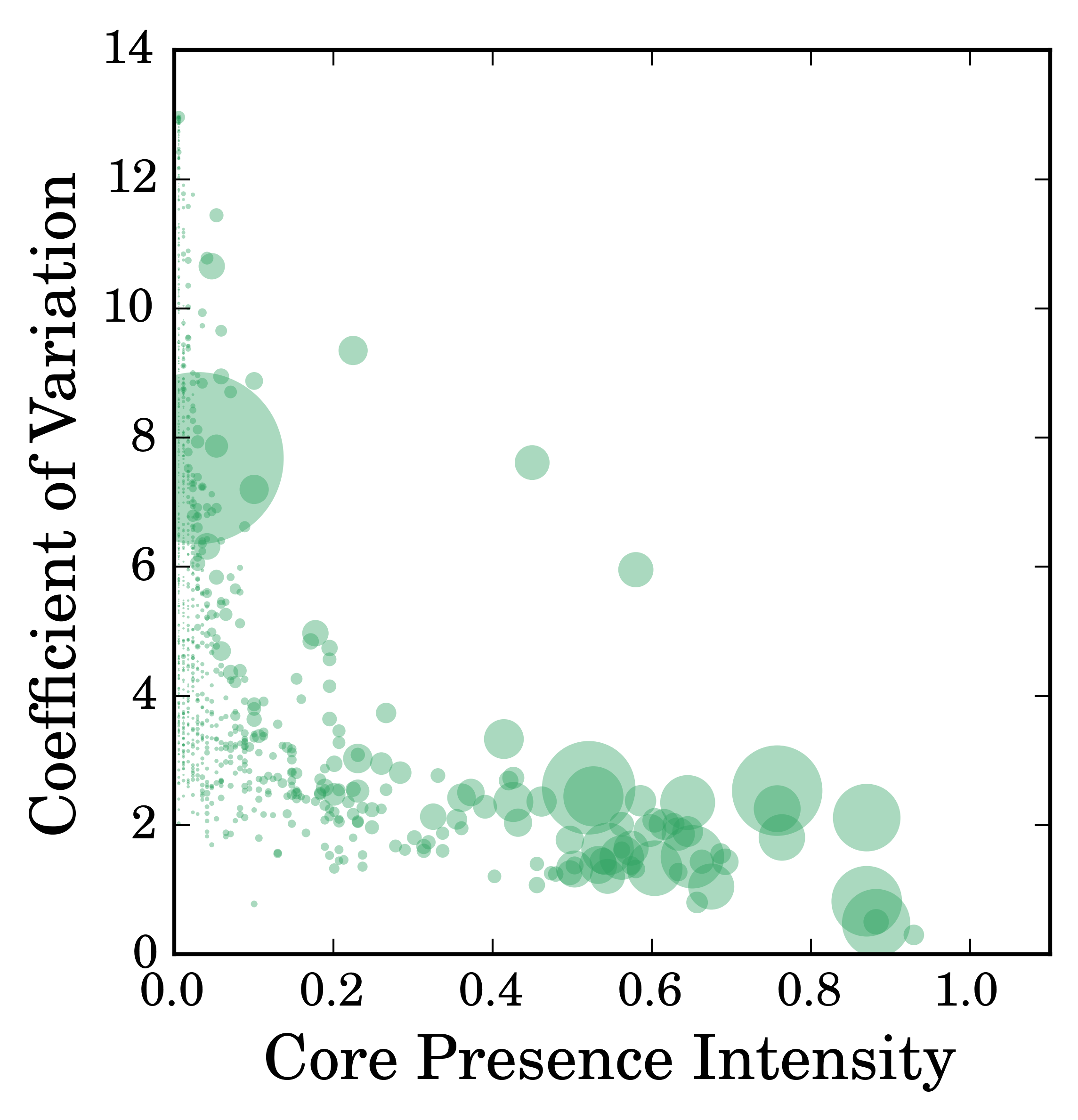}
                \caption{SC, 11065 prefixes}
                \label{fig:cv_cp_sc}
        \end{subfigure}
\caption{Relation between $I_{cp}$ and $c_v$}  
\label{fig:cv_cp}
\end{figure*}
Finally, it is interesting to study the correlation between these two metrics: $c_v$ 
and $I_{cp}$. 
They are plotted  together in Fig.~\ref{fig:cv_cp} for two representative sites (SA and SC)\footnote{A comprehensive set of graphics is available on a research report~\cite{RR}}, where each circle represents a prefix and its radius is proportional to the prefix in weekly volume share.
The biggest circles mostly concentrate in the lower right corner of each sub-graph, which corresponds to the remarks made previously.
However, some exceptions exist, especially on SC (but also SF, SG and SI), where we notice circles of big sizes having low $I_{cp}$ and relatively high $c_v$.
These prefixes bring significant traffic volume share within a short duration, which makes predictive prefix selection difficult. 
Finally, it's not a surprise to see that the $c_v$ of prefixes with big week volume is, to a certain extent, inversely correlated to their $I_{cp}$. 

\subsection{Quantitative index of traffic burstiness}

In order to define traffic burstiness more formally and to systematically estimate this property, we define 
the index $\beta(P)_h$, for a prefix $P$ and  a certain hour $h$ 
as:
$\beta(P)_h = 
         -\log(I_{cp}(P)) \times vp(P)_h$ 
         if $I_{cp}(P) > 0$ 
        and $0$ otherwise, 
where $vp(P)_h$ is the hourly volume percentage of prefix $P$ at hour $h$.
The logarithmic term
on $I_{cp}$ aims at amplifying the volume contribution from prefixes with rare \textit{core} presence, and attenuating the influence of prefixes being intensively in the \textit{core}.
A large $\beta$ value indicates that it is hard to predict the volume associated with this prefix while representing a significant hour volume.

In order to estimate the overall impact brought by bursty traffic at hour $h$, we sum up the $\beta(P)_h$ for each $P$ inside the \textit{core} of that hour: more formally, 
$BI_h = \sum_{P \in \textit{core}_h} \beta(P)_h$.
For all the networks, we estimate their traffic burstiness with the mean and maximum value of $BI$ series over the week. The results are given in Table~\ref{tab:bi}. The maximum $\beta$ over all the week is also given for the sake of clarity.
\begin{table}[!tb]
\centering
\begin{tabular}{cccc}\toprule
\textbf{Network} & \textbf{Mean $BI$} & \textbf{Max $BI$} & \textbf{Max $\beta$}\\
\midrule
SA & 14.61 & 37.79  &  7.44\\
SB & 31.09 & 46.85  &  4.08\\
SC & 40.57 & 145.07 &  145.05\\
SD & 42.14 & 69.34  &  18.10\\
SE & 20.91 & 44.30  &  20.17\\
SF & 44.10 & 98.69  &  78.77\\
SG & 51.21 & 125.41 &  102.05\\
SH & 15.91 & 35.06  &  16.29\\
SI & 38.59 & 85.47  &  56.56\\
\bottomrule
\end{tabular}
\caption{Traffic burstiness.}
\label{tab:bi}
\end{table}


We found that this simple metric is able to capture the burstiness of a given site. The existence of prefixes representing significant volumes but with fairly low $I_{cp}$ , such as for the site SC in Fig.~\ref{fig:cv_cp}, correspond to the biggest maximum $\beta$ values.
On the contrary, a small value of $BI$ (around 30 or lower) is arrived for sites such as SA, with more predictable volume-significant prefixes.

\begin{figure*}[!ht]
		\centering
        \begin{subfigure}[b]{0.49\textwidth}
                \centering                                        \includegraphics[width=0.66\textwidth]{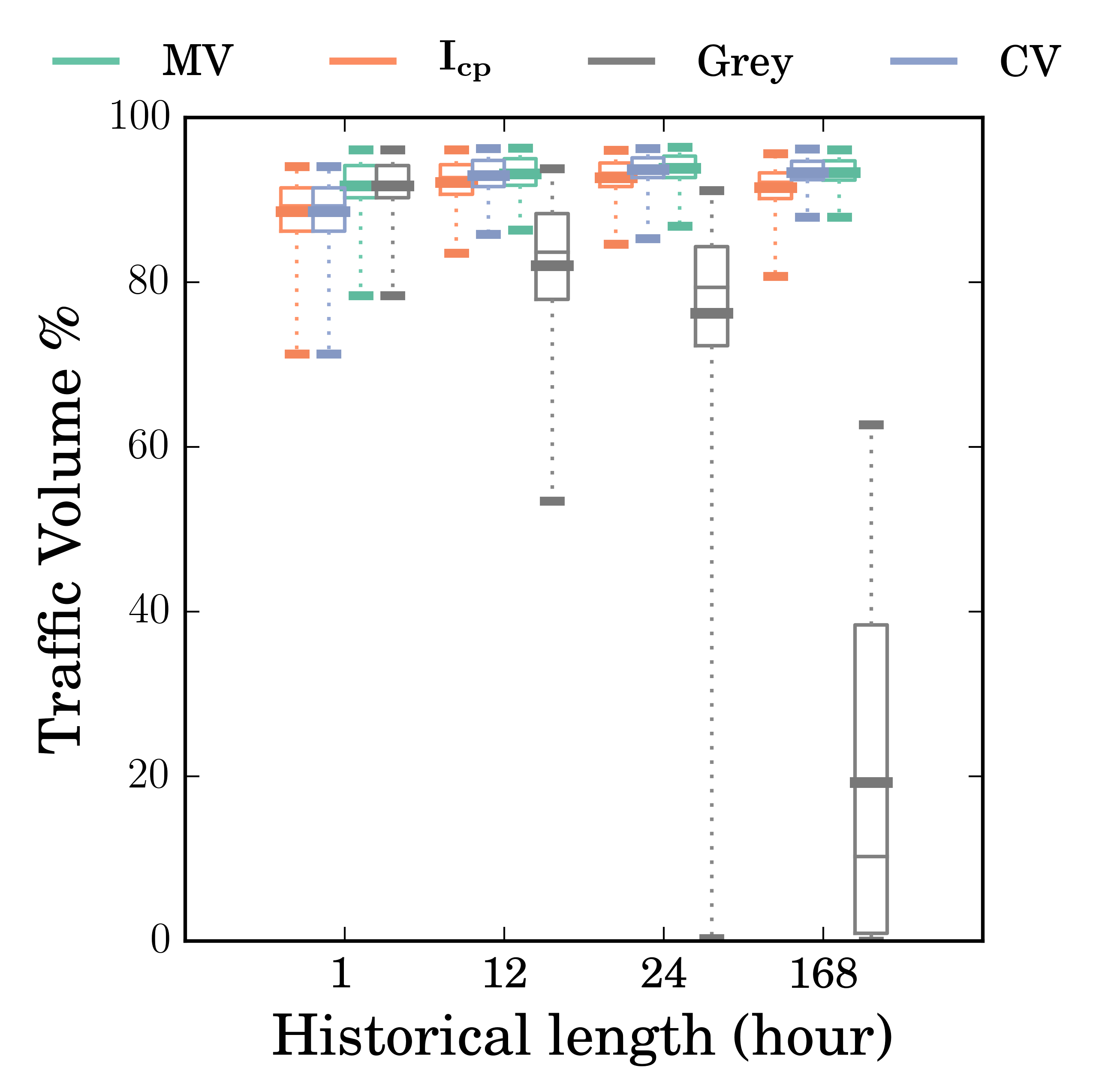}
                \caption{SA}
                \label{fig:cvg_sa}
        \end{subfigure}
        \hfill
        \begin{subfigure}[b]{0.49\textwidth}
                \centering                \includegraphics[width=0.66\textwidth]{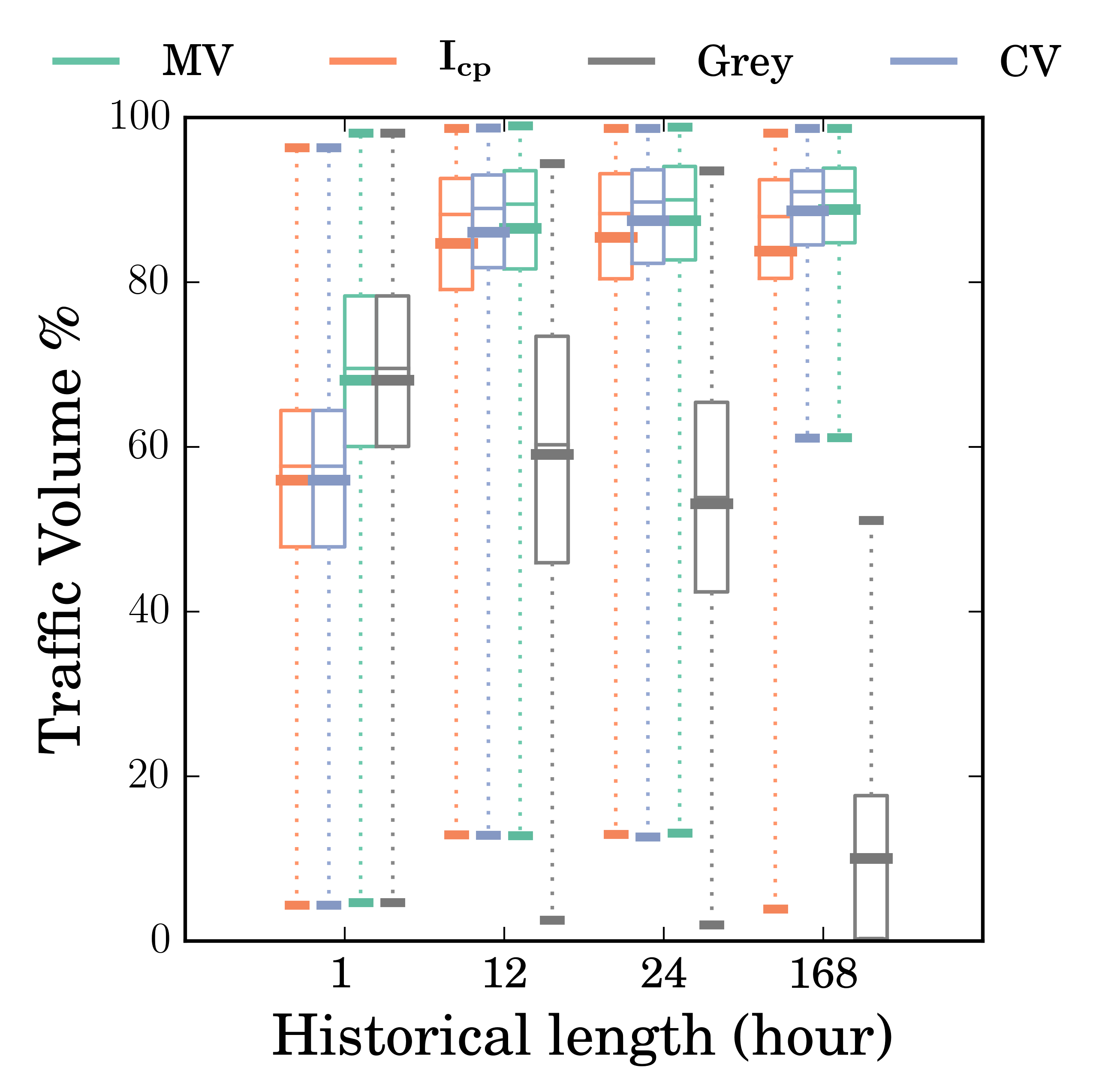}
                \caption{SC}
                \label{fig:cvg_sc}
        \end{subfigure}
\caption{Hourly volume fraction covered by the selected prefixe set}
\label{fig:cvg}
\end{figure*}

\section{Predictive prefix selection}
\label{sec:sele}
Based on the previous observations, several measures are possible for predicting the``importance'' of a prefix. They are enumerated below. 
\paragraph*{Mean Volume}
With this metric, we predict that at hour $h+1$, the volume importance of prefix $P$ is indicated by it's mean hourly volume over the last $L$ hours, $MV(P,L)_{h+1} = 1/L \times \sum_{i = h-L+1}^{h} v(P)_i$.
It is based on the observations from Fig.~\ref{fig:cv}, that top prefixes ranked over the week tend to have smaller hourly volume variation around their mean volume.
\paragraph*{Core Presence Intensity}
The prediction could use $I_{cp}(P,L)_{h+1} = 1/L \times \sum_{i = h-L+1}^{h} cp(P)_i$, i.e. the \textit{core} presence intensity of the prefix over the last $L$ hours. It derives from the observation made from Fig.~\ref{fig:cpi}, that high ranked prefixes by their week volume are more likely to have intense \textit{core} presence.

\paragraph*{Core Volume}
Finally, the prediction could be based on $CV(P,L)_{h+1} = 1/L \times \sum_{i = h-L+1}^{h} cp(P)_i \times v(P)_i$,  a combination of both $MV$ and $I_{cp}$.
$CV$ has the potential to be more resource thrifty compared to $MV$, as it is be calculated only for those prefixes which ever appeared in the \textit{core} over the last $L$ hours --- while $MV$ is computed for all active prefixes. From our observations, the \textit{core} size actually represents $5\%$ to $50\%$ of all active prefixes.

In previous work by Zhange et al.\ \cite{Zhang2012} on FIB caching, a grey differential model $GM(1,1)$ \cite{Julong1989} is employed to  predict which BGP prefixes will represent the  biggest packet counts. 
For the sake of comparison, we implemented the $GM(1,1)$ model as well in our work. 



\begin{figure*}[!tb]
\centering
		\centering
        \begin{subfigure}[b]{0.49\textwidth}
        \centering
                \includegraphics[width=0.66\textwidth]{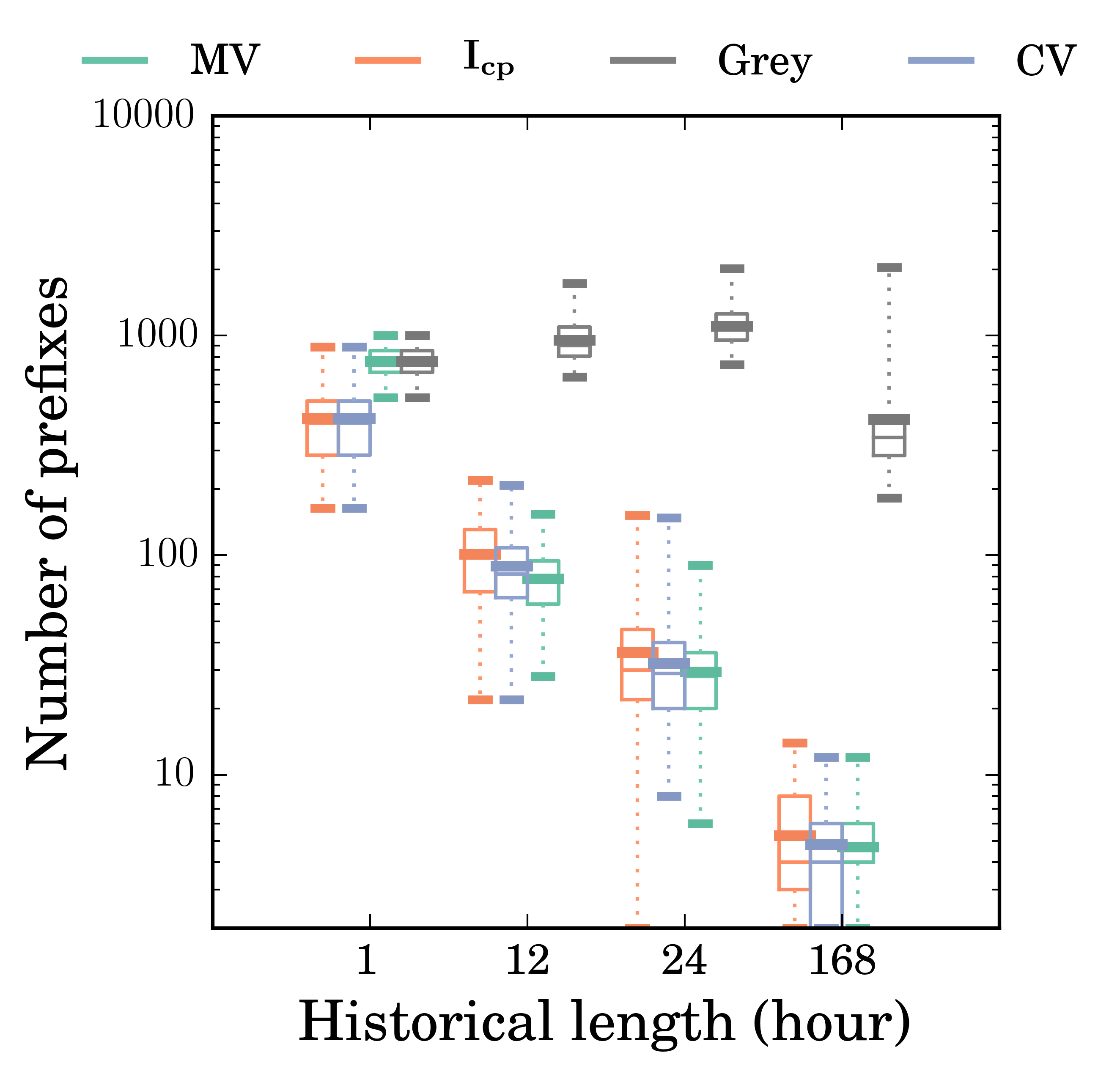}
                \caption{SA}
                \label{fig:churn_sa}
        \end{subfigure}
        \hfill
        \begin{subfigure}[b]{0.49\textwidth}
        \centering
                \includegraphics[width=0.66\textwidth]{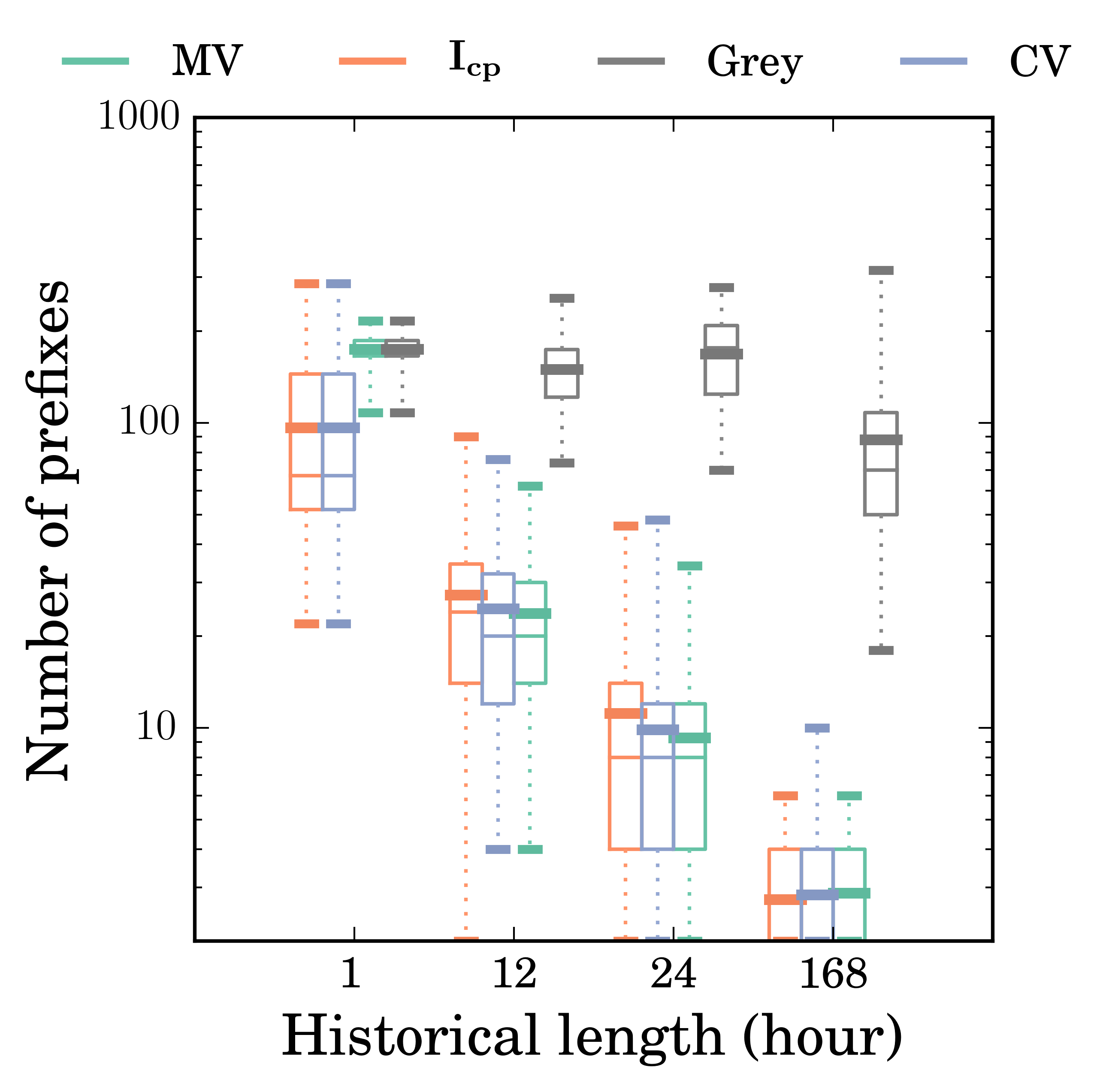}
                \caption{SC}
                \label{fig:churn_sc}
        \end{subfigure}
\caption{Hourly churn of the selected prefix set}
\label{fig:churn}
\end{figure*}

We fix the selection size to the maximum \textit{core} size over the week. 
We can then evaluate the quality of the prefix set selected with the different proposed metrics, or with the $GM(1,1)$ model. 

Fig.~\ref{fig:cvg} illustrates the hourly volume coverage by the four methods in the form of boxplot, representing the minimum, maximum, 25th and 75th percentile, medium and mean values. 
Among proposed metrics, we find that the $CV$ is very close to $MV$ in terms of volume coverage, proving that it is a good approximation of the later, while being more resource economical.

Basing solely on the last 1 hour records, all methods yield already a mean volume coverage $>80\%$ on SA, SB, SE and SH, which implies a strong continuity on prefix volumes between two consecutive hours. 
On SC and SG, however, using records of last 168 hours, i.e. a week, offers much better minimum volume coverage than shorter records. This is due to the fact that at certain hour, SC and SG undergo a great amount of bursty traffic (as observed previously, e.g. in Table~\ref{tab:bi} and Fig.~\ref{fig:cv_cp}). By increasing the historical length, the selection metrics are able to have better visibility into the past and capture some of these bursty prefixes ---  finally improving the minimum coverage.
This gain in minimum volume coverage by using long historical records can actually be observed on all networks, between 1 hour and 12 hours, also between 12 hour and 24 hour. 
However from 24 hour to 168, this gain doesn't necessary happen on all sites, which is due to the fact that the total volume brought by some highly bursty prefix is diluted by the long time span using $MV$ and $CV$ metrics. In order to capture them, a larger selection set size is need, which inevitably includes more prefixes of few significance.

On the other hand, the hourly volume coverage by grey model drops as we increase the historical records length and is in general much worse than the metrics proposed in this work. 
In order to understand the underlying reason, we used $GM(1,1)$ model described above to dynamically predict the total hourly volume of all prefixes, which is normally much more regular and smoother than the volume series of individual prefixes. It sorts out that the grey model reacts to variations in a delayed manner when $L$ being big and extravagantly overreacts to sudden changes when $L$ is small~\cite{RR}. 





Fig.~\ref{fig:churn} gives the results concerning the churn of the selection prefix set (with a boxplot representation).
The churn is defined as the difference (i.e. number of new and deleted prefixes) between the new predicted set and the previous one. A high churn could pose operational problems: when probing is considered, for instance,  a stable set is clearly more adapted. Sarrar et al.\ \cite{Sarrar2012} also argued that small prefix churn is very important in network architecture with decoupled forwarding and control  planes, such as SDN (Software Defined Networking), as it leads to lower communication overhead.

As expected, a clear drop of the churn value can be seen when the historical length increases --- as opposed to what happens with the grey model. 
In that sense, using long historical records can be a wise choice in practical uses.
Furthermore, for networks with relatively few bursty traffic, e.g. SA, the mean volume coverage with last 168 hour records is extremely close to that with last 24 hour records, as seen in Fig.~\ref{fig:cvg_sa}.
Finally, for networks with highly bursty traffic, SC, using long records has the potential to obviously improve worst-case volume coverage. 

In the purpose of lowering churn, Sarrar et al.\ \cite{Sarrar2012} proposed selecting top prefixes over time bins of different lengths (ranging from 1 second to 10 minute in their FIB-caching environment). In our context, we found that the difference in mean volume coverage using record lengths larger than 1 hour is marginal, thus little gain can be expected from this method. 



\begin{figure}[!tb]
\centering
		\centering
        \begin{subfigure}[b]{0.21\textwidth}
        \centering
                \includegraphics[width=\textwidth]{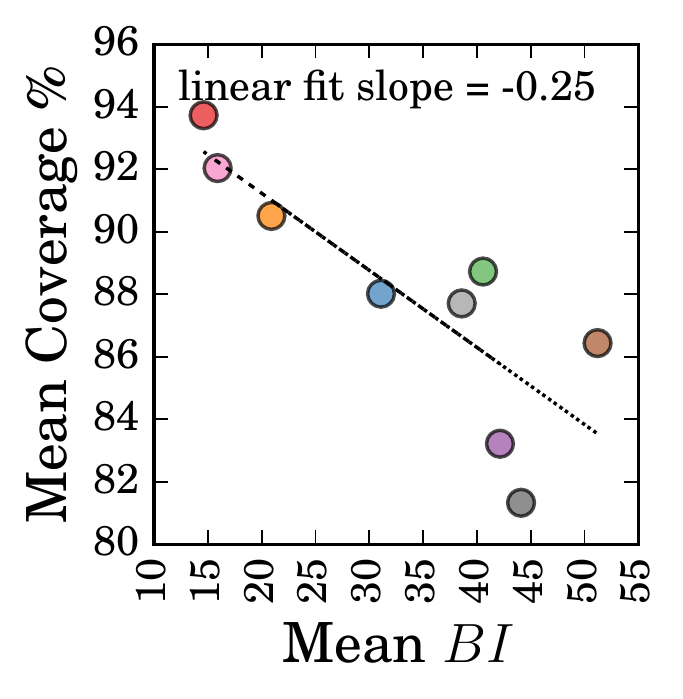}
                \caption{Average level}
                \label{fig:bi_cvg_mean}
        \end{subfigure}
        \hfill
        \begin{subfigure}[b]{0.27\textwidth}
                \includegraphics[width=\textwidth]{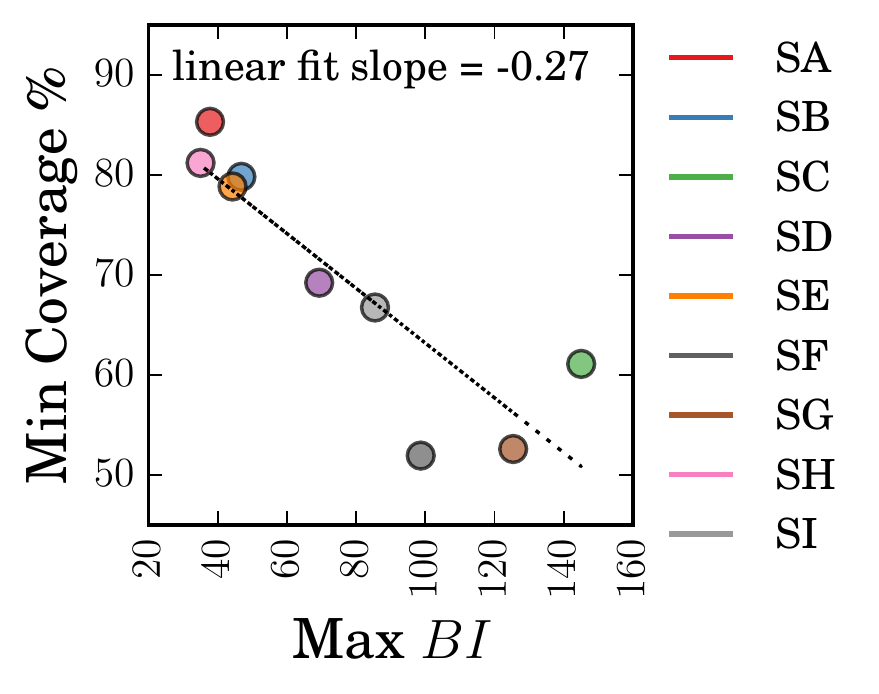}
                \caption{Worst case}
                \label{fig:bi_cvg_worst}
        \end{subfigure}
\caption{The relationship between burstiness index $BI$ and traffic volume coverage of selected prefix using $CV$ metric.}
\label{fig:bi_cvg}
\end{figure}

\begin{figure*}[!tb]
\centering
		\centering
        \begin{subfigure}[b]{0.49\textwidth}
        \centering
                \includegraphics[width=0.6\textwidth]{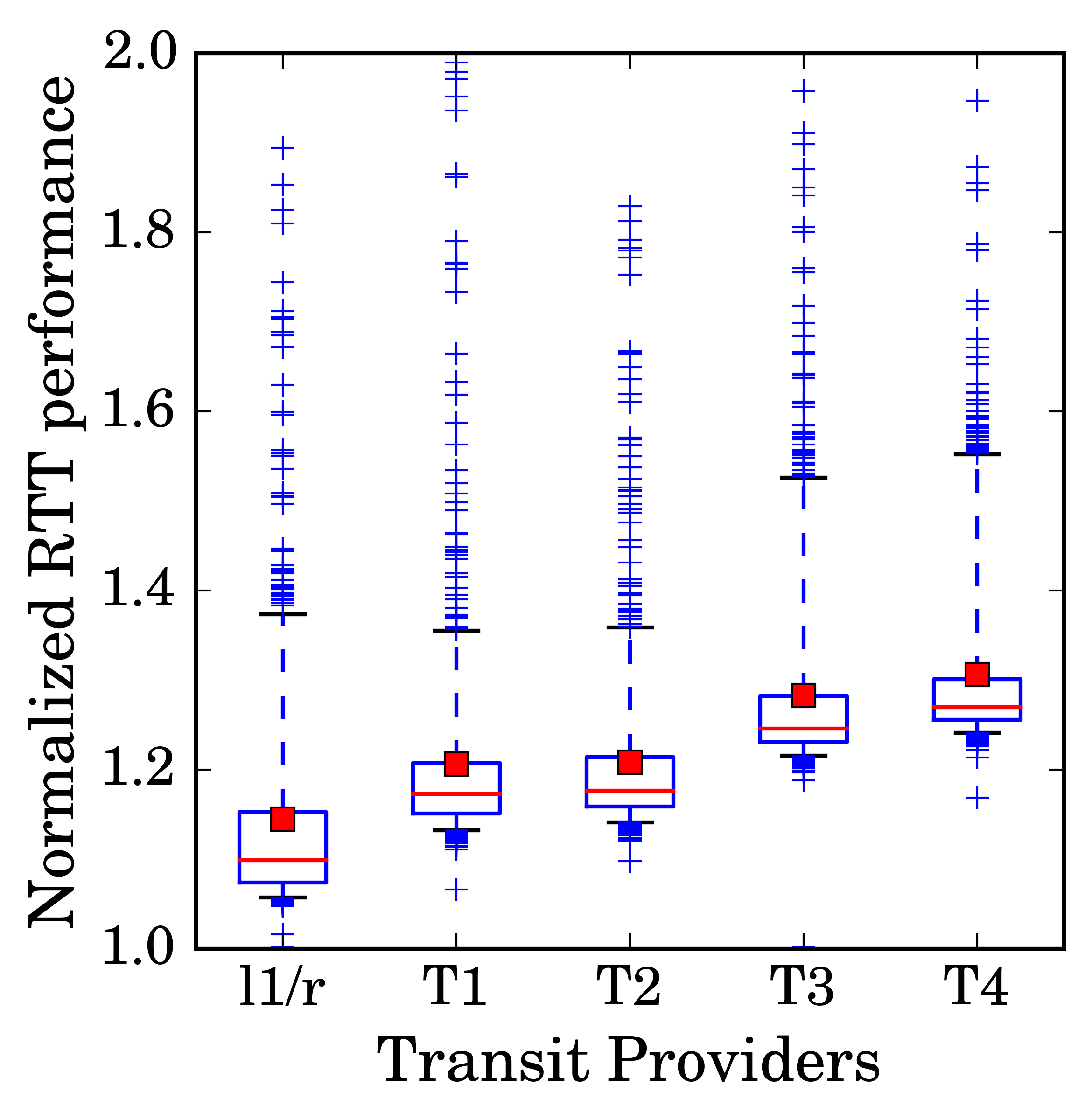}
                \caption{SA}
                \label{fig:np_sa}
        \end{subfigure}
        \hfill
        \begin{subfigure}[b]{0.49\textwidth}
        \centering
                \includegraphics[width=0.6\textwidth]{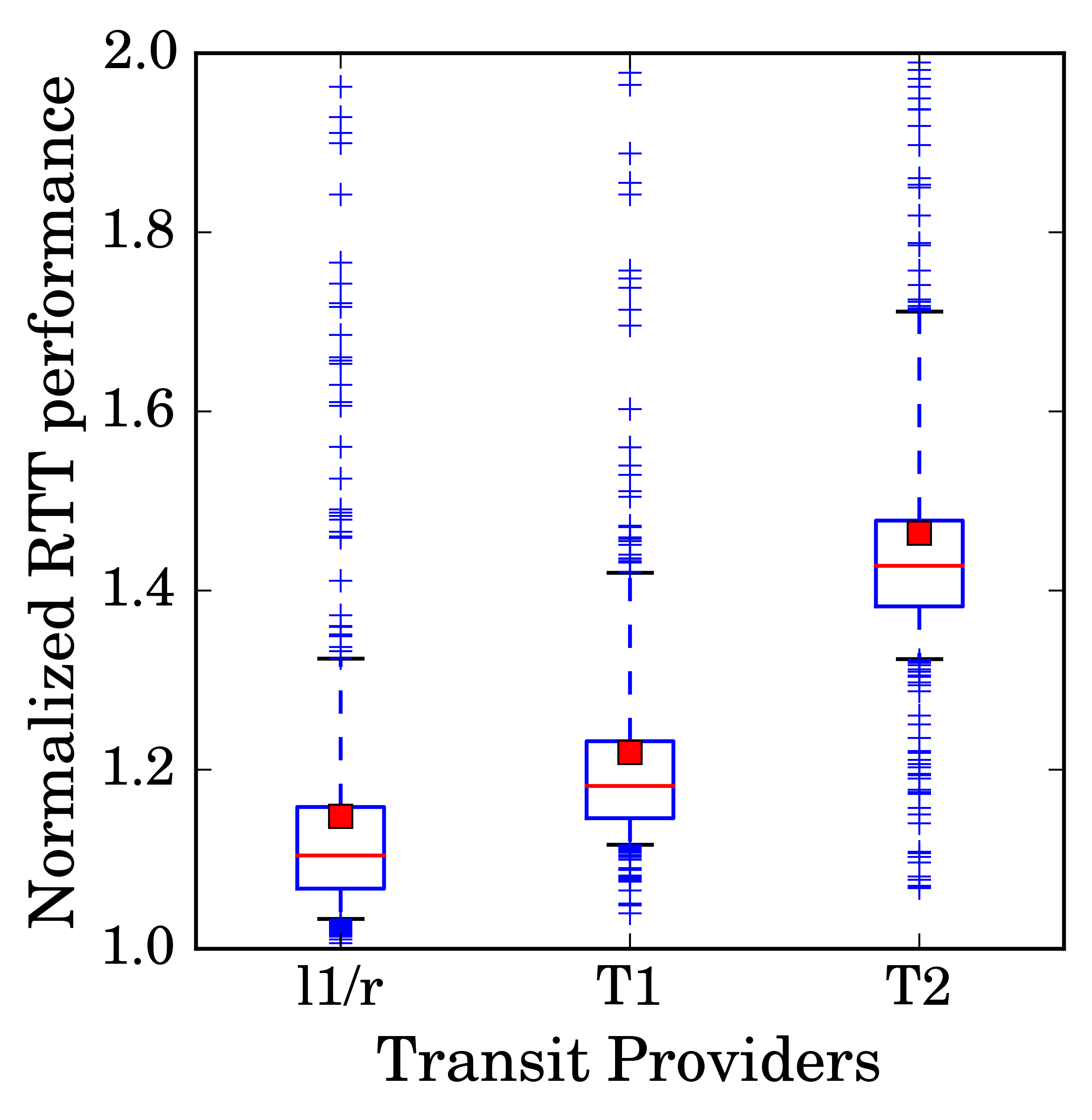}
                \caption{SC}
                \label{fig:np_sc}
        \end{subfigure}

\caption{Normalized RTT performance with active probing.} 
\label{fig:np}
\end{figure*}

Finally, in Fig.~\ref{fig:bi_cvg} the mean/minimum coverage achieved with $CV$ metric is showed as a function of the $BI$ index. We can see that the mean (resp. minimum) coverage is inversely proportional to the mean (resp. maximum) $BI$ index. As expected, this graphic highlight the difficulty to cover a large trafic volume for networks with more bursty traffic. However, the quasi-linear curve obtained proves that $BI$ is a very meaningful metric to identify sites with bursty trafic. For large $BI$ values, it can be worthy of choosing a larger prefix set size (which was previously fixed to the weekly maximum core size), if possible.

\section{RTT probing and transit performance evaluation}
\label{sec:rtt}

We then study the potential performance that our multi-homing optimization system can gain. We first present the method used to evaluate the transit performance in terms of RTT (Round-Trip Time) perceived by selected prefixes.


TCP SYN scanning was employed to measure RTT between selected prefixes and the local network.
The probe traffic toward a selected prefix was steered on all available transit providers, by means of explicit routing.
For a pair of selected prefix and transit provider, a probe is scheduled at 240 second interval in average with $30\%$ randomization left or right. 
The probe data covers the entire week starting from June 1st.


Our performance evaluation method follows the same principle as the one proposed by Akella et al.\ \cite{Akella2003a}: 
The computation is done as :
\begin{align}
NP^{Tx}_{t_i} = \frac{1}{|SP|} \sum_{P \in SP} M^{Tx}_{t_i}(P)/\min_{Tj \in T}M^{Tj}_{t_i}(P)
\label{eq:np}
\end{align}
where $NP^{Tx}_{t_i}$ is the normalized RTT performance for transit provider $Tx$ at probe tickle $t_i$, and 
$M^{Tx}_{t_i}(P)$ denotes the RTT measured toward selected prefix $P$.

This measure is first normalized over the best measure among all available transit providers $T$ at the same probe tickle.
Then we average this normalized RTT over all prefixes inside the selected prefix set $SP$.
If a transit provider offers the smallest RTT to all selected prefixes, it should have a normalized RTT performance equaling to 1.

Fig.~\ref{fig:np} gives the results of transit performance evaluation for SA, in a ``boxplot'' format, where the intervals represent $5^{th}$ and $95^{th}$ percentiles (values below and beyond, marked by a \texttt{+} symbol, can be regarded as outliers).
Along the X-axis, available transit providers are aligned increasingly according to their mean normalized RTT performance (marked by a square). 

$l1/r$ denotes a virtual transit provider simulated by a dynamic route selection algorithm inspired from \cite{Akella2008}. The method selects the transit provider that provides the smallest RTT in last round of probing for each selected prefix. This choice is  based on the hypothesis that RTT of a path demonstrates temporal locality and thus is  closely related to its most recent measurement.
It is a basic algorithm used for the sake of illustration only.

The number of prefixes probed on each network and traffic volume represented are given in Fig.~\ref{fig:cvg_sa}. 

We observe that the performance differences among different transit providers is particularly clear, in particular on SC.
This confirms that multi-homing can still provide significant performance improvement nowadays. 
However this gain in performance is not inherently given in the context of BGP.
Even for transit provider that offers the best best mean $NP$, there exist moments where its performance deviates far above 1, which means that traffic toward some selected prefixes are suffering from RTTs much larger than on other available transit providers.
This implies that a network will not have optimal  performance if it uses transit providers in a static and indistinguishable manner for each individual prefixes.
Therefore a dynamic route selection  method (such as $l1/r$) that dynamically select the best outgoing next-hop for each individual selected prefix is needed. 
$l1/r$ selects the transit that provides the smallest RTT in last probing round. 
In the case that measurement data is not available, it chooses randomly. 
For all networks except SG, the simulated algorithm out-performs all physical transit providers.
On SB, the overall RTT is $20\%$ below the mean RTT of the best available transit provider.
Still the $NP$ value of this virtual provider varise within a wide range, which calls for further investigation on the characters of RTT variation in time.

\section{Related work}
\label{sec:bg}
Some previous works \cite{Feamster2003, Akella2008, Goldenberg2004} acknowledged the importance of preforming inter-domain TE only for destinations that matter.
However, no general solution was proposed to predictively select the BGP prefixes with important traffic volume.
Some other works leveraged the skewed distribution of Internet traffic to downsize FIB \cite{Iannone2007, Ballani2009, Kim2009, Zhang2012, Sarrar2012, Liu2015}, where understanding of traffic dynamism on much shorter time scales, e.g. seconds and minutes, are required. 
In our work, we used traffic traces of coarser time resolution over longer period of time --- conditions more adapted to the complicated operations associated required for inter-domain TE.


Zhang et al.\ \cite{Zhang2012} assumed that the stability and popularity of Internet traffic is positively correlated without verification. 
Papagiannaki et al.\ \cite{Papagiannaki2004} showed that this correlation doesn't necessarily exist for 5-tuple flows on 5 minute interval. 


Akella et al.\ \cite{Akella2003a} quantified the performance gain from multi-homing using traces from a large CDN network more than ten years ago. 
In a latter work \cite{Akella2008}, they evaluated a dynamic route selection system on a testbed, but with only 100 destinations. 

\section{Conclusion and future directions}
\label{sec:fut}
The present paper tackled the problem of controlling the majority of data traffic via 
a small subset of BGP prefixes, by exploiting the uneven Internet traffic distribution. 
The main goal being the simplification of management procedures associated with inter-domain Traffic Engineering, in particular for multi-homed sites.  
One of the challenges in addressing such problem was to select in a scalable manner the prefixes that will carry most traffic volume in the forthcoming time.

We analyzed real traffic measurements from nine different sites (located in five different countries) to understand the distribution of traffic volume associated with different BGP prefixes, as well as its variation in time.   
We observed how the most important prefixes (representing largest volume over a week) are generally stable in time, with small hourly variations around their mean. 
Based on our observations, we proposed three simple 
metrics (also easy to compute) to proactively select prefixes with important foreseeable traffic volume.
We demonstrated that the metrics we proposed lead to better volume coverage compared to the existing solutions.
Furthermore, we measured RTT performance through different transit providers and simulated a route decision engine on the selected prefix set. 
Through such analysis we showed that  with such dynamic route selection the overall RTT performance of the network can be improved by $20\%$ compared to the best available transit provider. 

In order to achieve better prefix selection methods, we have shown that capturing bursty prefixes is the key point. 
Interesting directions to explore in order to improve the already good results we achieved is to group prefixes by their activity profiles. 
For each group, selection method could be adapted to the its traffic dynamism.
When dealing with prefixes with regular volume patterns and small hourly variation, the simple metrics proposed in the work perform extremely well and may suffice. 
Nevertheless, for bursty prefixes, we might need a more sophisticate model that extracts additional activity features, like for instance long term periodicity.
The use of the burstiness index $\beta$, shown to be very expressive, in prefix classification is also worthy of further work.

\bibliographystyle{IEEEtran}
\bibliography{IEEEabrv,ref}
\end{document}